\documentclass[conference]{IEEEtran}
\IEEEoverridecommandlockouts
\usepackage{cite}
\usepackage{amsmath,amssymb,amsfonts}
\usepackage{algorithmic}
\usepackage{graphicx}
\usepackage{textcomp}
\usepackage{xcolor}
\usepackage{authblk}
\def\BibTeX{{\rm B\kern-.05em{\sc i\kern-.025em b}\kern-.08em
    T\kern-.1667em\lower.7ex\hbox{E}\kern-.125emX}}
\begin{document}

\title{Automatic spinal curvature measurement on ultrasound spine images using Faster R-CNN}

\author[1]{Zhichao Liu}
\author[1]{Liyue Qian}
\author[1]{Wenke Jing}
\author[1]{Desen Zhou}
\author[1]{Xuming He}
\author[2]{Edmond Lou}
\author[1,3,*]{Rui Zheng}
\affil[1]{\small School of Information Science and Technology, ShanghaiTech University, Shanghai, China}
\affil[2]{\small Electrical \& Computer Engineering Dept., University of Alberta, Canada}
\affil[3]{\small Shanghai Engineering Research Center of Intelligent Vision and Imaging, ShanghaiTech University, Shanghai, China}

\maketitle

\begin{abstract}
Ultrasound spine imaging technique has been applied to the assessment of spine deformity. However, manual measurements of scoliotic angles on ultrasound images are time-consuming and heavily rely on raters’ experience. The objectives of this study are to construct a fully automatic framework based on Faster R-CNN for detecting vertebral lamina and to measure the fitting spinal curves from the detected lamina pairs. The framework consisted of two closely linked modules: 1) the lamina detector for identifying and locating each lamina pairs on ultrasound coronal images, and 2) the spinal curvature estimator for calculating the scoliotic angles based on the chain of detected lamina. Two hundred ultrasound images obtained from AIS patients were identified and used for the training and evaluation of the proposed method. The experimental results showed the 76.1\% AP on the test set, and the Mean Absolute Difference (MAD) between automatic and manual measurement was 4.3° which was within the clinical acceptance error (5°). Meanwhile the correlation between automatic measurement and Cobb angle from radiographs was 0.79. The results revealed that our proposed technique could provide accurate and reliable automatic curvature measurements on ultrasound spine images for spine deformities.
\end{abstract}

\begin{IEEEkeywords}
Ultrasound spine imaging, Spinal curvature, Faster R-CNN, Scoliosis
\end{IEEEkeywords}

\section{Introduction}
Adolescent idiopathic scoliosis (AIS) is a 3D deformity of spine featured with lateral deviation and axial vertebral rotation which occurs in the general population ranging from 0.93 to 12\% \cite{b1}, and a population-based, cross-sectional study showed that the prevalence rate of idiopathic scoliosis is 5.14\%. According to Fan et al \cite{b2}, the AIS patients can benefit from preventive treatment significantly. Nowadays, Cobb angle measured on the standing posterior-anterior radiographs is the gold standard to diagnose and monitor AIS. Cobb angle is defined as the angle between the endplates of the most titled vertebra of the spine \cite{b3}. However, AIS patients require long-term examinations to monitor changes in spine curves, which can result in high exposure to radiation during childhood.

\par

In recent years, Ultrasound (US) Imaging has been developed and studied by many scholars because of its characteristics of non-radiation, flexible operation and low cost, and it becomes one of the most promising methods in scoliosis research \cite{b4}. Zheng et al.\cite{b5} studied the intra- and inter-rater reliabilities of measuring coronal curvatures obtained from ultrasound images using the center of lamina(COL) method, and validated that the difference between US and radiographic measurements was within the range of clinical acceptance error (5°)\cite{b3}. They also found that with the aid of the previous radiographs, the reliability and accuracy of US measurement had been improved significantly \cite{b6}.

\par

Some automatic measurement methods for spine curves have been presented. Zeng et al.\cite{b7} leveraged the gradient vector flow(GVF) snake model to automatically locate the spinous process(SP) to evaluate the curve severity of scoliosis. Zhou et al.\cite{b8} proposed a new automatic measurement of spinal curvature on VPI image by extracting vertebral feature including transverse and spinous processes using two-fold threshold and phase congruency. Nowadays, the deep learning methods especially the Convolutional Neural Networks (CNN) have been applied for the feature detection and assessment of spine and vertebra morphological characteristics. Wu et al. \cite{b9} proposed a new method based on MVC-Net to implement the fully automatic end-to-end measurement of Cobb angles on radiographs. The proposed network took anterior-posterior and lateral radiographs as input and detected the joint features in multiple views. The features were then used to locate the landmarks on the spine radiograph, and finally the Cobb angles were estimated by utilizing the dependence between joint features and landmarks. Ungi et al. trained the U-net\cite{b10} to accomplish the bone structure segmentation on ultrasound spine sagittal images which was then used for 3-dimensional volume reconstruction to accomplish the spine curve measurement, and the spine curves were manually measured on the  ultrasound reconstructed image. The results showed that the average difference in scoliosis measurement between ultrasound and X-ray was 2° \cite{b11}.

\par

The objectives of this study are to apply deep learning method for fully automatic measurement of spinal curvature and to reduce the subjective influence from the manual measurement.

\section{Method}

\subsection{Framework for Automatic Measurement of Spinal Curvature}

\begin{figure*}[htbp]
\centerline{\includegraphics[width=0.7\textwidth]{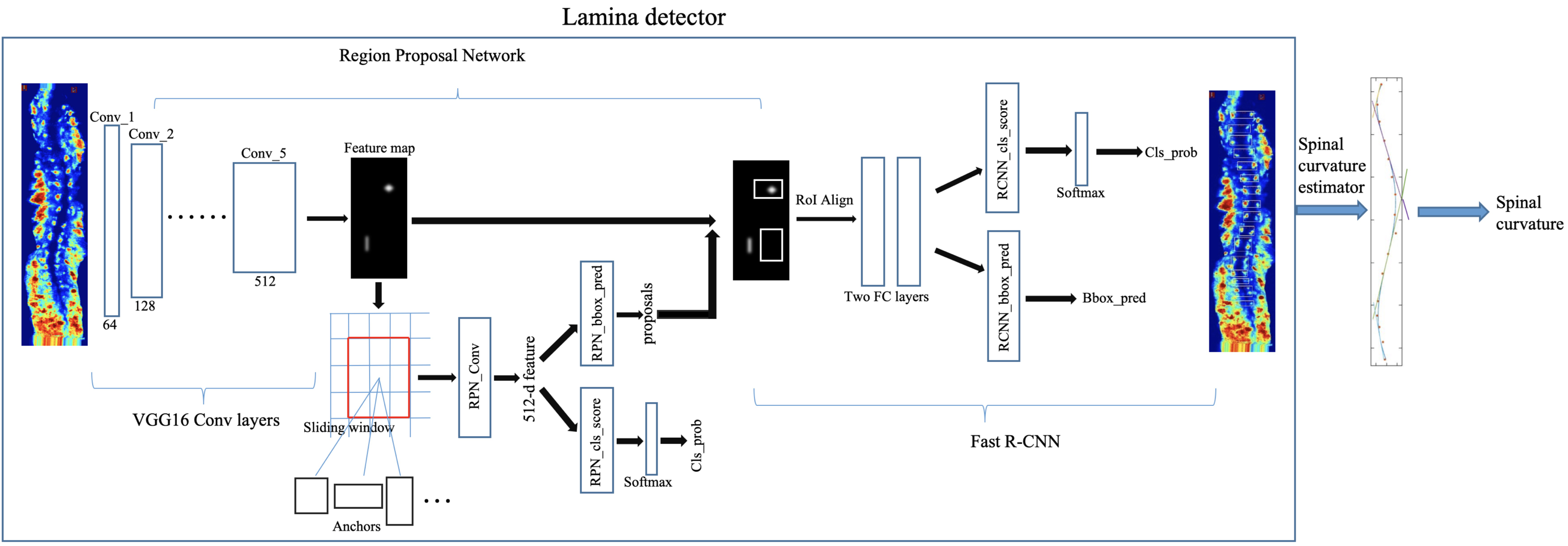}}
\caption{The framework for automatic measurement of spinal curvature. The first module was lamina detector base on Faster R-CNN, the output detection boxes were fed into spinal curvature estimator for final spinal curvature results.}
\label{fig1}
\end{figure*}

In this study, we attempted to apply the deep learning method to automatically detect the vertebral features and measure proxy Cobb angle i.e. scoliotic curvatures on ultrasound spine images. In comparison to radiography, ultrasound images showed more drawbacks and deflects due to the imaging mechanism including low signal-to-noise ratio (SNR),unclear structures and boundaries. We proposed a new framework for automatic measurement of spinal curvature through lamina detection using Faster R-CNN \cite{b12} and global inference for spine curves based on the centers of detected lamina boxes. The framework consisted of two closely-linked components as shown in Fig. \ref{fig1}: (1) the lamina detector which detected the lamina structures on the input ultrasound coronal spine image, and (2) the spinal curvature estimator which leveraged the detected lamina structure and obtained the spine curves by describing the overall bending trend of spine.

\subsubsection{Lamina Detector}
Lamina detector based on Faster R-CNN was a coarse-to-fine pipeline(two stages) to accomplish the task of lamina detection. For the first stage,  the coronal ultrasound image was fed into the feature extractor which extracted the semantic information and provided them to Region Proposal Network  (RPN) to output relatively coarse detection results called proposal. For the second stage, the fine detector (Fast R-CNN)\cite{b13} took in the proposal boxes and corresponding proposal features and then output the final detection result of lamina pair.

\begin{itemize}
    \item Region Proposal Network: RPN was a fully convolutional network and consisted of a $3\times3$ convolutional layer to process feature map and two $1\times1$ convolutional layer for classification and regression respectively. RPN performed sliding window on 512-d feature map and output 512-d processed feature which was fed into regression layer to compute proposals. With the sliding window mechanism, RPN ensured the leverage of the whole feature map. In addition, RPN acquired thee multi-size detection boxes by introducing the Anchor of  different size, which included three scales $\{128^{2},256^{2},512^{2}\}$ and three ratios $\{0.5, 1, 2\}$ in our case. The generated anchors could be considered as the initial detection box, and through the coarse-to-fine process, a more accurate lamina detection box was obtained.
    \item Fast R-CNN: Fast R-CNN took the proposals generated by RPN as input and then performed further classification and bounding box regression to obtain more accurate lamina detection results. Since the Fast R-CNN was composed of a fully connected layer, the size of proposals was required to be same. In fact, the size of proposals was not consistent. Therefore, our lamina detector adopted RoI Align\cite{b14} to resize the proposals and avoided round errors occurring in ROI Pooling\cite{b12}, which could ensure the spatial accuracy.
\end{itemize}

\subsubsection{Spinal curvature estimator}
\begin{figure}[htbp]
\centerline{\includegraphics[width=0.4\textwidth]{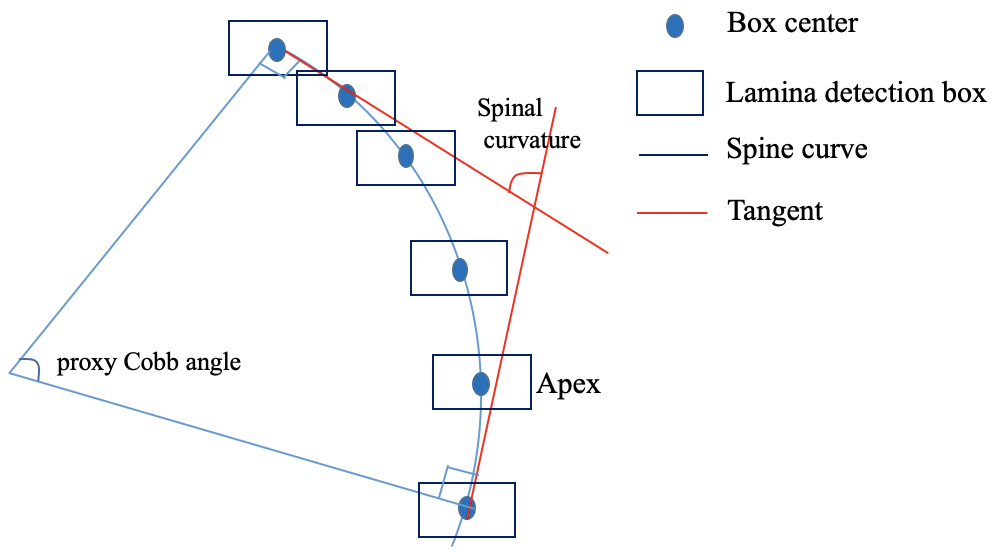}}
\caption{Automatic measurement schematic. We obtained the box centers, fitted the curve that describing the spine bending trend by $5^{th}$ polynomials and found two points with largest change of the derivative values to the apex, finally the angle between the tangent lines at the two points was spinal curvature.}
\label{fig2}
\end{figure}

The assumption of automatic  measurement for the spinal curvature was that the center of lamina detection box represented the position of vertebra and the fitting curve by connecting all centers was approximate to spine curves, therefore the angle derived from the curve was the proxy Cobb angle as called spinal curvature. The calculation process was shown in Fig. \ref{fig2}, and specific steps were as follows:
\begin{itemize}
    \item Extracting the centers of lamina detection boxes and obtaining the spine curves by $5^{th}$ polynomial fitting.
    \item Locating the apex points of the curve, and dividing the curve into two sections with the apex as cut-off point.
    \item Finding the two points with the largest change of the derivative values relative to apex, and calculating the angle formed by the two tangent lines at these two points which was called spinal curvature.
\end{itemize}

\subsection{Dataset}

The ultrasound image data sets were collected from untreated AIS patients aged from 11 to 16 years. The inclusion criteria were as follows: (1) without surgical treatment; (2) the US scan and posterior-anterior standing radiograph obtained within one hour; (3) The Cobb angle ranges from 10° to 45°. All patients participating in the experiment signed a consent form in advance and the ethical approval was granted by the local health research ethics committee. All US scans were obtained using SonixTABLET system equipped with 128-element C5-2/60 GPS transducer and SonixGPS system(Analogic Ultrasound - BK Medical, Peabody, Massachusetts, USA). The US transverse frames combined with position information were acquired and then  fed into the reconstruction algorithm. After that, the 3D volumes were projected onto coronal plane to obtain coronal image\cite{b4}. Two hundred reconstructed ultrasound coronal images were used for the training and validation process, and each image contained 14-17 pairs of laminae. We randomly divided all images into 3 sections: training set (140 images), validation set (30 images), test set (30 images). The brightness and contrast of all images were adjusted to highlight the laminae pair, and some horizontal flip were applied in the preprocess for the data enhancement.

\subsection{Evaluation metrics}

The overall framework was implemented based on the Pytorch deep learning framework using Python. Experiments were carried out with one NVIDIA Geforce RTX 2070 on Linux server. To evaluate the performance of Lamina detector, average precision (AP) of IoU=0.5 was applied.

\par

For the spinal curvature estimator, we evaluated the performance using mean absolute difference (MAD), standard deviation (SD) and the correlation coefficient (R) between our automatic measurement and Cobb method/ ultrasound manual measurement. With one rater with 10 years experience, these results are convincing. The Cobb angles of all participated AIS subjects were acquired from clinical records and used as the gold standard to evaluate the curvature measurement from ultrasound data. For the US manual measurement, the COL method [5]was applied to measure the spinal curvature by one rater with 10 years ultrasound experience.

\section{EXPERIMENTAL RESULTS}

\begin{figure}[htbp]
\centerline{\includegraphics[width=0.35\textwidth]{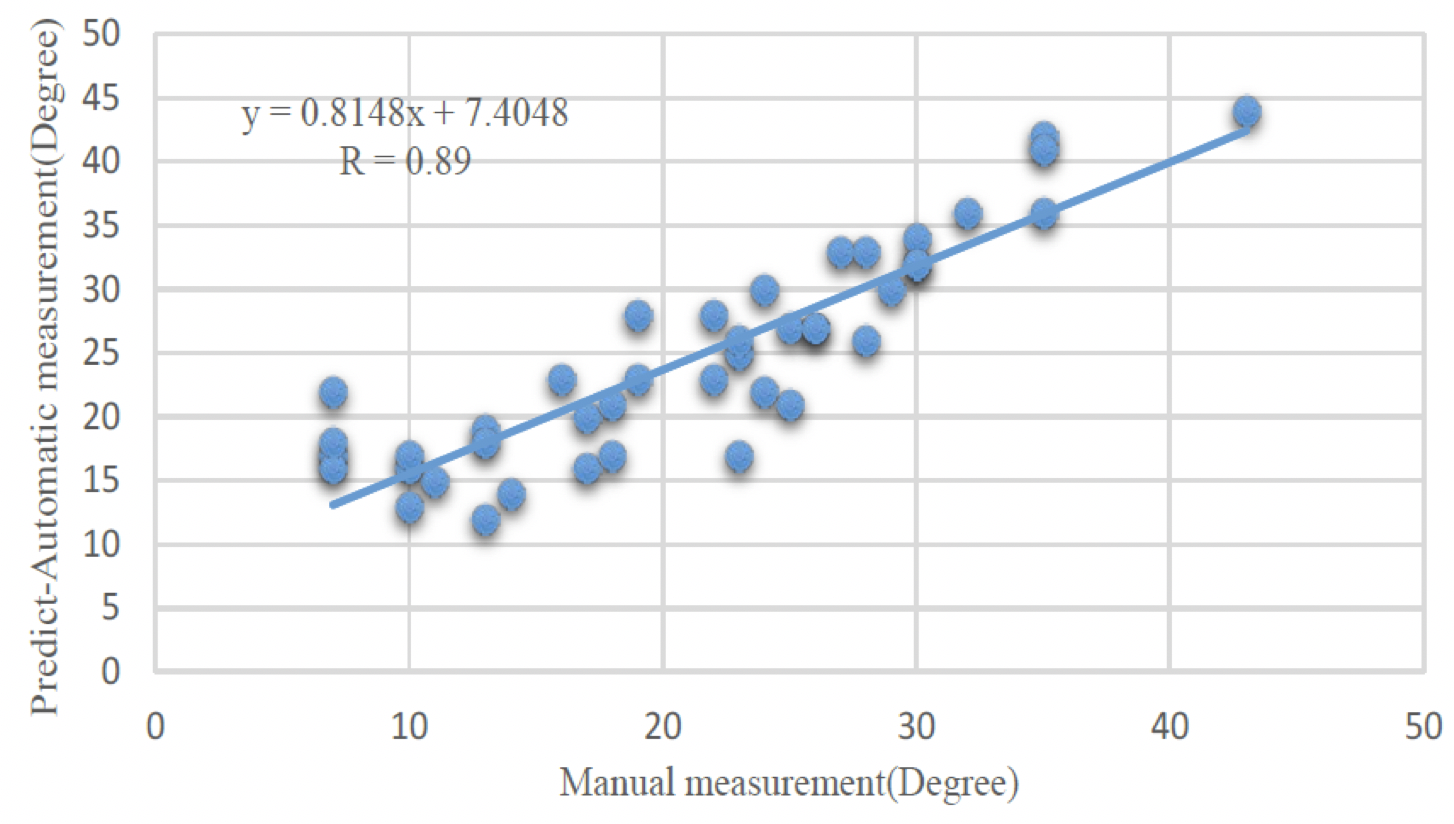}}
\caption{The correlation between the automatic measurement and US manual measurement.}
\label{fig3}
\end{figure}

\begin{figure}[htbp]
\centerline{\includegraphics[width=0.35\textwidth]{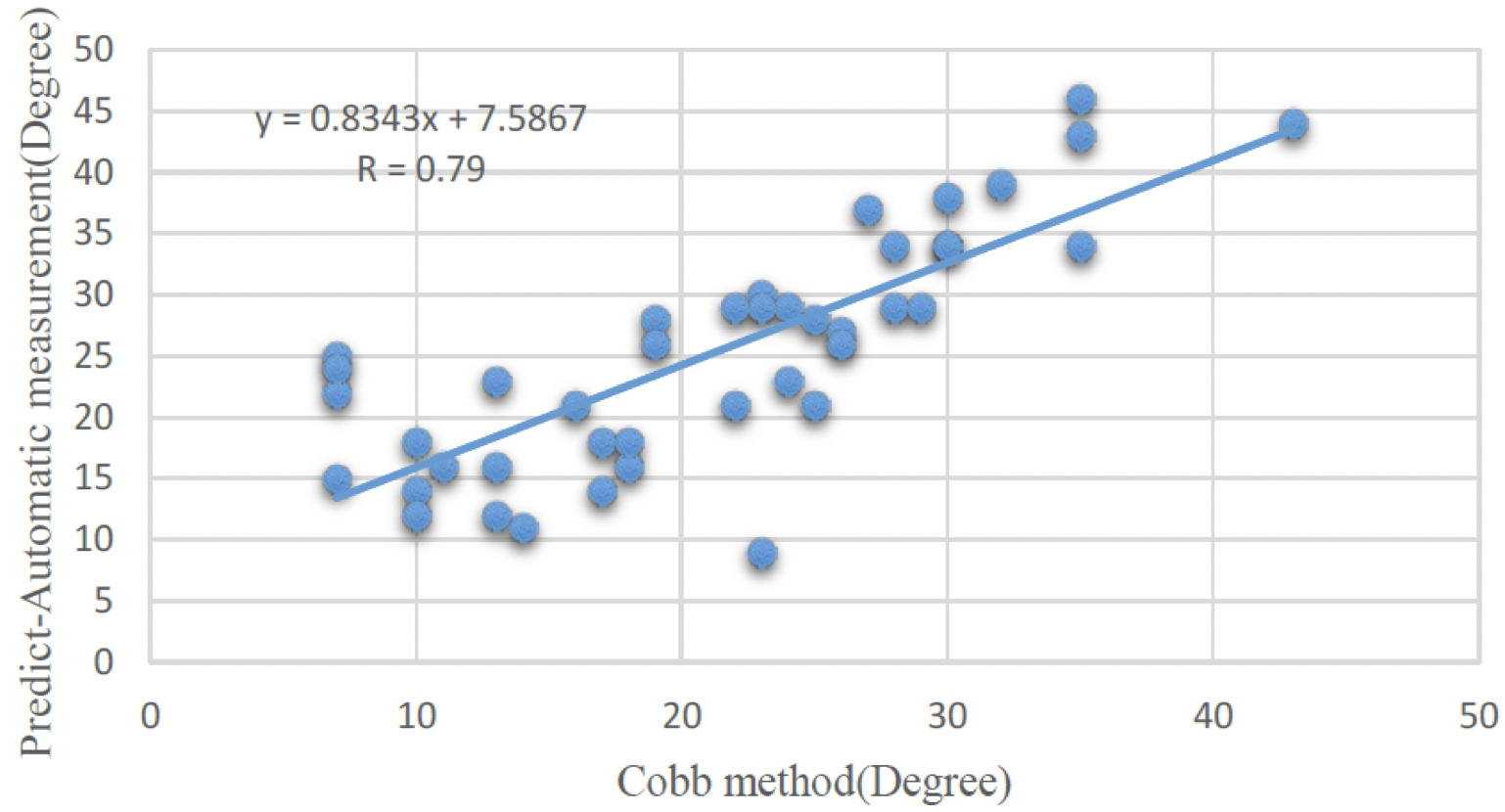}}
\caption{The correlation between the automatic measurement and X-ray Cobb measurement.}
\label{fig4}
\end{figure}

\begin{figure}[htbp]
\centerline{\includegraphics[width=0.35\textwidth]{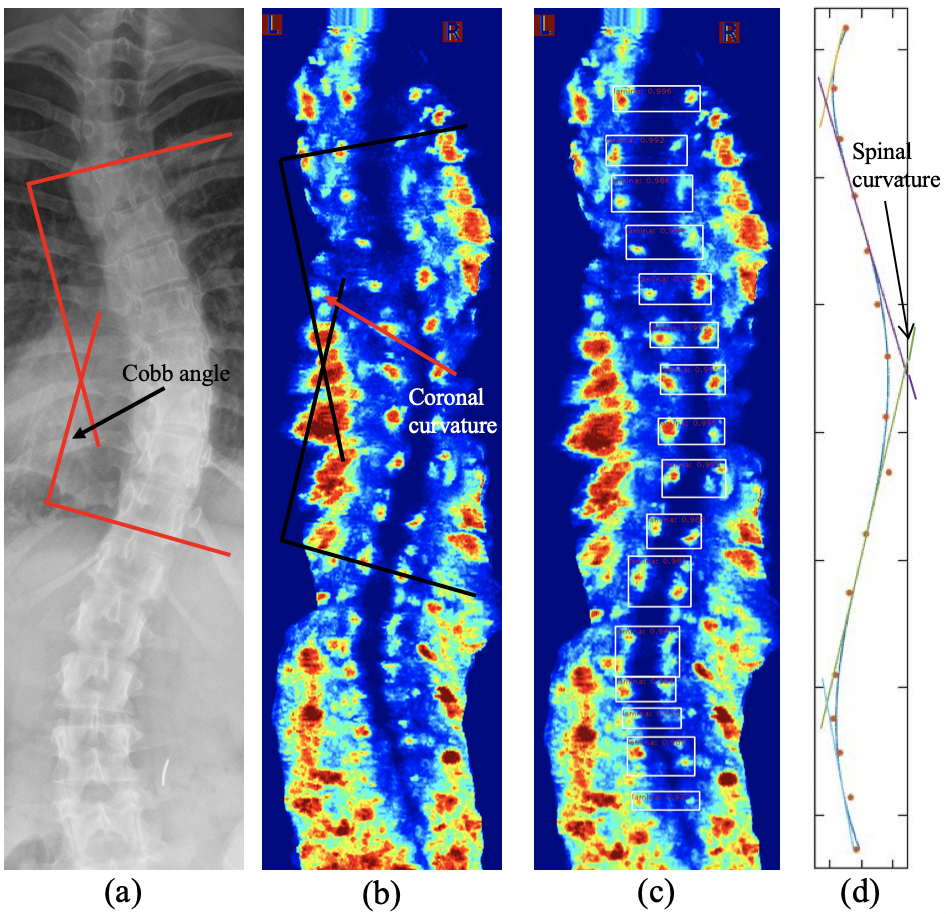}}
\caption{The comparison of curve measurement for different modalities: (a) radiographic Cobb angle, (b) US manual coronal curvature, (c) auto detection of lamina on US coronal images and (d) US automatic spinal curvature.}
\label{fig5}
\end{figure}

The AP for lamina detector was 76.1\%. The results of spinal curvature estimation for 30 test data sets were shown in  Table \ref{tab1}. We compared the automatic ultrasound measurement results with the manual measurement from the ultrasound and Cobb angles from X-ray images, and the mean absolute differences were 4.3° and 5.5° respectively. Fig. \ref{fig3} \& \ref{fig4} illustrated the correlation among different measurement modalities, and the correlation coefficients were 0.89 and 0.79 respectively. Fig. \ref{fig5} illustrated an example of ultrasound automatic curve measurement in comparison to the radiographic and US manual results.

\begin{table}[htbp]
\caption{THE STATISTICS OF CURVATURE MEASUREMENT}
\begin{center}
\begin{tabular}{|c|c|c|c|}
\hline
\multicolumn{2}{|c|}{Automatic vs US manual} & \multicolumn{2}{c|}{Automatic vs Cobb} \\ \hline
$MAD\pm SD$                 & $R$                   & $MAD\pm SD$               & $R$               \\ \hline
$4.3\pm 3.3^{\circ}$       & $0.89$                & $5.5\pm 4.6^{\circ}$      & $0.79$              \\ \hline
\end{tabular}
\label{tab1}
\end{center}
\end{table}

\section{DISCUSSION}

\begin{figure}[htbp]
\centerline{\includegraphics[width=0.48\textwidth]{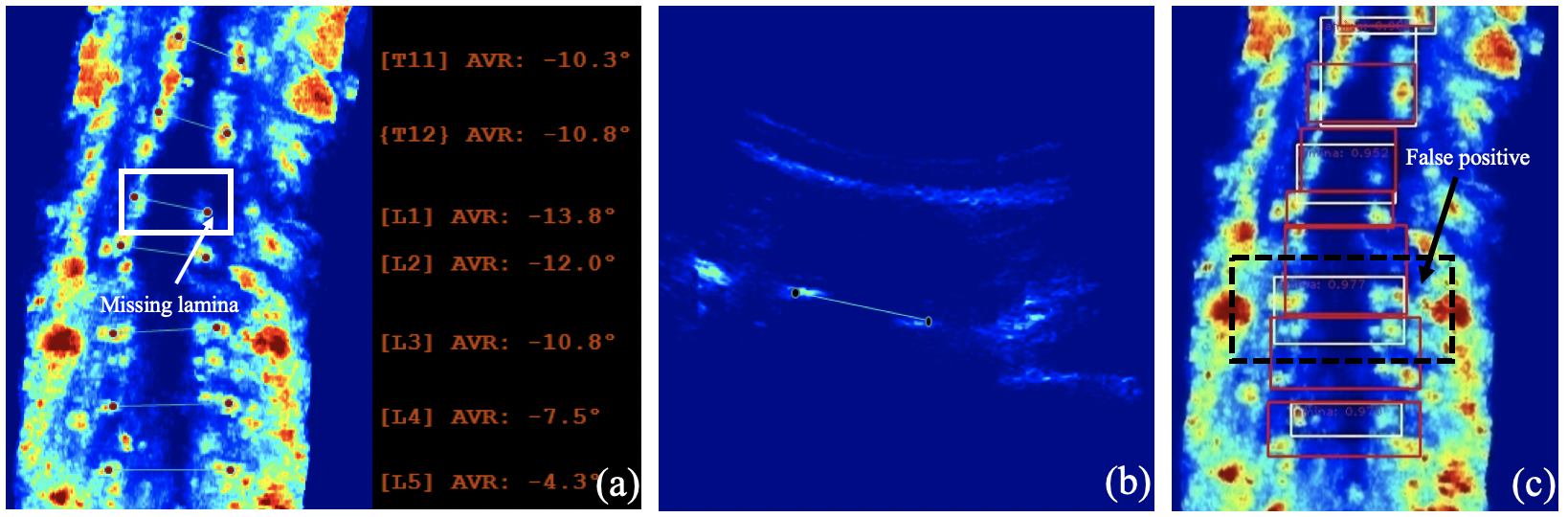}}
\caption{The images from a patient with largest AVR of 13.8°. (a) Coronal ultrasound image. (b) Transverse image. (c) Comparison of detection results (white box) and ground truth lamina box (red box).}
\label{fig6}
\end{figure}

The axial vertebral rotation (AVR) often exists on the deformed spine and leads to incomplete and unclear lamina structure on ultrasonic images. It is because the titled spinous process blocks part of ultrasonic signals toward lamina that causes the feature loss, for example the missing lamina on the right side illustrated in the white box in Fig. \ref{fig6}a. Fig. \ref{fig6}b showed the transverse image with a large AVR of 13.8°, and the left side of lamina with high intensity was much easier to identify than the right side. In the case of incompleteness and dimness of lamina structure, the lamina detector would inevitably produce some false positives (FPs) as shown in black dotted box of Fig. \ref{fig6}c, the red boxes represented the ground truth and the white boxes represented the detection boxes. In the US manual measurement process, the FPs will bring the inevitable interference due to the incorrect detected lamina.

\par

\begin{figure}[htbp]
\centerline{\includegraphics[width=0.3\textwidth]{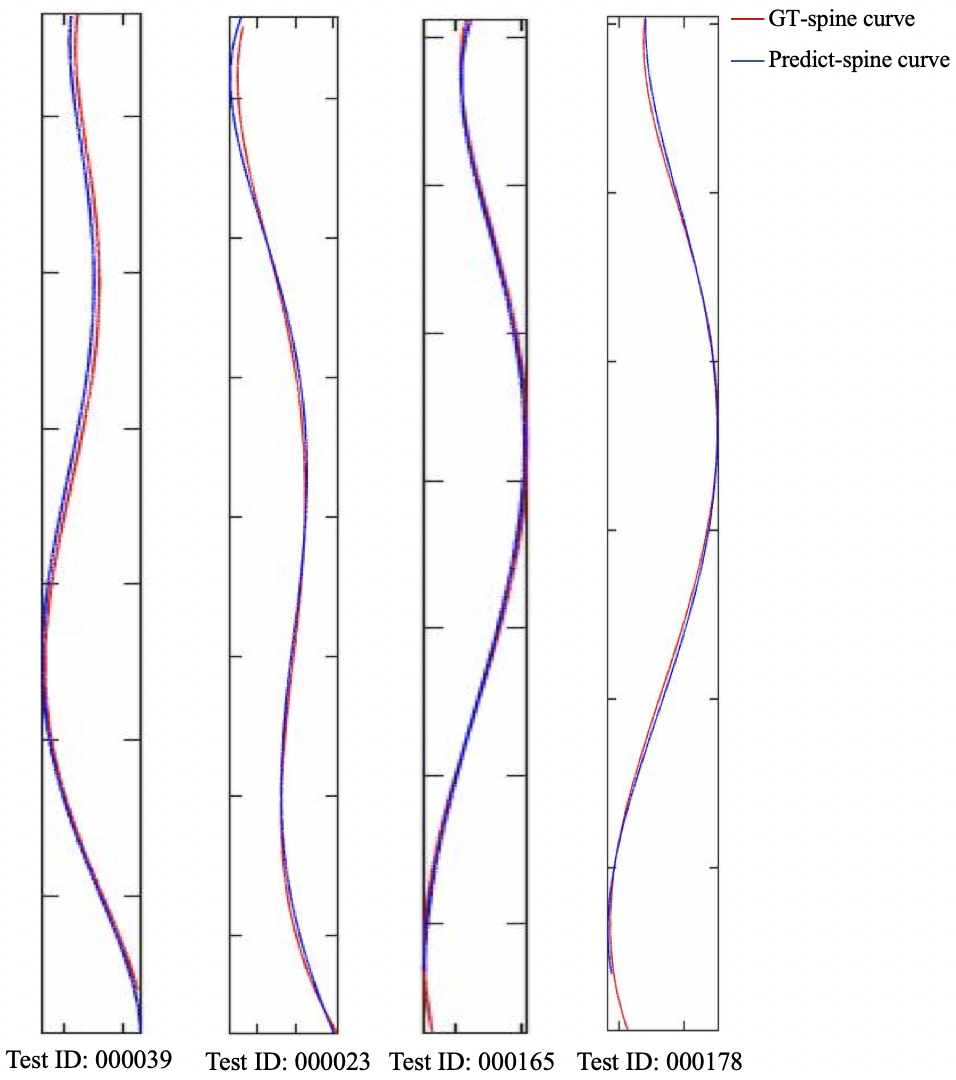}}
\caption{The comparison between ground truth (GT)-spine curve and Predict-spine curve. The matching results indicated that the negative impact of FPs could be well reduced by the global inference on spine curve.}
\label{fig7}
\end{figure}

The physiological structure of spine such as ribs and lamina adjacency with each other provide a good prior to detect the spinal structures, However, the lamina detector implicitly learned these prior which could sometimes lead to false positive with high scores in the middle part of spine, and cause the center point of FPs not properly deviating from the path of spine curve. Thus, we implemented global inference on spine curve using polynomial fitting process to reduce the negative impact of FPs. AS is shown in Fig. \ref{fig7}, the curves from the detected results were well overlapped with the curve from the ground truth, and it indicated that global inference could describe the overall bending trend of spine.

\section{CONCLUSION}

In this paper, we proposed a fully automatic measurement framework based on faster R-CNN method for spinal curvature  assessment of AIS.  The experimental results showed that the automatic measurement  result was comparable with radiographic Cobb and US manual methods, and the measurement differences were within the range of clinical acceptance error. Moreover, the global inference from the curve fitting process based on spine physiological structure could reduce the influence of FPs. Therefore, the proposed framework could provide a feasible, accurate and promising modality to assess spine deformity using ultrasound imaging technique. In the future, we’ll focus on improving the image equality and locating the center point of lamina with keypoints detection algorithm to follow the process of US manual measurements, i.e. to acquire the proxy Cobb angle.

\section*{Acknowledgment}
This work was sponsored by Natural Science Foundation of Shanghai (Grant No.19ZR1433800).




\begin{thebibliography}{00}
\bibitem{b1} Negrini S, Donzelli S, Aulisa A G, et al. 2016 SOSORT guidelines: orthopaedic and rehabilitation treatment of idiopathic scoliosis during growth[J]. Scoliosis and spinal disorders, 2018, 13(1): 1-48.
\bibitem{b2} Hengwei Fan, Zifang Huang, Qifei Wang, et al. Prevalence of idiopathic scoliosis in Chinese schoolchildren: a large, population-based study[J]. Spine, 2016, 41(3): 259-264.
\bibitem{b3} J. R. Cobb, “Outline for the study of scoliosis,” The American Academy of Orthopedic Surgeons Instructional Course Lectures, 1948, 5, 261-275.
\bibitem{b4} Chen H, Zheng R, Qian L, et al. Improvement of 3D Ultrasound Spine Imaging technique using Fast Reconstruction Algorithm[J]. IEEE Transactions on Ultrasonics, Ferroelectrics, and Frequency Control, early access, 2021.
\bibitem{b5} Zheng R, Chan A C Y, Chen W, et al. Intra-and inter-rater reliability of coronal curvature measurement for adolescent idiopathic scoliosis using ultrasonic imaging method—a pilot study[J]. Spine Deformity, 2015, 3(2): 151-158.
\bibitem{b6} Zheng R, Young M, Hill D, et al. Improvement on the accuracy and reliability of ultrasound coronal curvature measurement on adolescent idiopathic scoliosis with the aid of previous radiographs[J]. Spine, 2016, 41(5): 404-411.
\bibitem{b7} Zeng H Y, Lou E, Ge S H, et al. Automatic Detection and Measurement of Spinous Process Curve on Clinical Ultrasound Spine Images[J]. IEEE Transactions on Ultrasonics, Ferroelectrics, and Frequency Control, 2020, 68(5): 1696-1706.
\bibitem{b8} Zhou G Q, Jiang W W, Lai K L, et al. Automatic measurement of spine curvature on 3-D ultrasound volume projection image with phase features[J]. IEEE transactions on medical imaging, 2017, 36(6): 1250-1262.
\bibitem{b9} Wu H, Bailey C, Rasoulinejad P, et al. Automated comprehensive adolescent idiopathic scoliosis assessment using MVC-Net[J]. Medical image analysis, 2018, 48: 1-11.
\bibitem{b10} Ronneberger O, Fischer P, Brox T. U-net: Convolutional networks for biomedical image segmentation[C]//International Conference on Medical image computing and computer-assisted intervention. Springer, Cham, 2015: 234-241.
\bibitem{b11} Ungi T, Greer H, Sunderland K R, et al. Automatic spine ultrasound segmentation for scoliosis visualization and measurement[J]. IEEE Transactions on Biomedical Engineering, 2020, 67(11): 3234-3241.
\bibitem{b12} Ren S, He K, Girshick R, et al. Faster r-cnn: Towards real-time object detection with region proposal networks[J]. Advances in neural information processing systems, 2015, 28: 91-99.
\bibitem{b13} Girshick R. Fast r-cnn[C]//Proceedings of the IEEE international conference on computer vision. 2015: 1440-1448.
\bibitem{b14} He K, Gkioxari G, Dollár P, et al. Mask r-cnn[C]//Proceedings of the IEEE international conference on computer vision. 2017: 2961-2969.
\end{thebibliography}
\end{document}